\documentstyle[preprint,aps]{revtex}
\begin{document}
\draft
\tighten

\title{Vacuum mimicking phenomena in neutrino oscillations}

\author{Osamu Yasuda\footnote{Email: yasuda@phys.metro-u.ac.jp}}

\address{Department of Physics, Tokyo Metropolitan University \\
Minami-Osawa, Hachioji, Tokyo 192-0397, Japan}

\date{June, 2001}
\preprint{
\parbox{5cm}{
hep-ph/0106232\\
}}

\maketitle

\begin{abstract}

It is shown generally that any oscillation probability in matter
with approximately constant density coincides with
that in vacuum to the first two nontrivial
orders in $\Delta m^2_{jk}L/E$ if $|\Delta m^2_{jk}L/E|\ll 1$ and
$|G_F N_e L|\ll 1$ are satisfied.

\end{abstract}
\vskip 0.1cm
\newpage

Recently a lot of efforts have been made on study of neutrino
oscillations at long baseline experiments.  Using the mass
hierarchical condition $|\Delta m^2_{21}|\ll|\Delta
m^2_{32}|\simeq|\Delta m^2_{31}|$ in the three flavor framework of
neutrino oscillations, it has been found in the case of T-conserving
probability $P(\nu_e\rightarrow\nu_\mu)$
\cite{DeRujula:1999hd,Freund:2000gy,Lipari:2001ds} or in the
case of T-violating probability $P(\nu_\mu\rightarrow\nu_e)$
\cite{Minakata:2000ee,Minakata:2001wm} that the oscillation probability
$P(\nu_\alpha\rightarrow\nu_\beta)_{\mbox{\rm\scriptsize matter}}$ in matter
coincides with that
$P(\nu_\alpha\rightarrow\nu_\beta)_{\mbox{\rm\scriptsize vacuum}}$ in vacuum
\begin{eqnarray}
P(\nu_\alpha\rightarrow\nu_\beta)_{\mbox{\rm\scriptsize matter}}\simeq
P(\nu_\alpha\rightarrow\nu_\beta)_{\mbox{\rm\scriptsize vacuum}}
\label{eqn:equality}
\end{eqnarray}
when $|\Delta m^2_{jk}L/E|\ll 1$ and $|A L|\ll 1$ are satisfied, where
$A\equiv\sqrt{2}G_F N_e$ stands for the matter
effect \cite{Mikheev:1986wj,Wolfenstein:1978ue} and
$N_e$ is the density of electrons.  This phenomenon was referred to as
vacuum mimicking in \cite{Minakata:2001wm}.
In this short note it is shown that
(\ref{eqn:equality}) holds in the first two nontrivial orders in
$\Delta m^2_{jk}L/2E$ and $AL$ (the terms quadratic and cubic in
$\Delta m^2_{jk}L/2E$ correspond to T-conserving and T-violating
probabilities in the leading order, respectively) for arbitrary
numbers $N$ of neutrino flavors with general form
diag$(A_1,A_2,\cdots,A_N)$ of the matter effect if $|\Delta
m_{jk}L/2E|\ll 1$ and $|A L|\ll 1$ are satisfied.

In the three flavor framework of neutrino oscillations,
the positive energy part of the Dirac equation which
describes neutrino propagation is given by
\begin{eqnarray}
i{d\Psi \over dt}=
\left[U{\mbox{\rm diag}}\left(E_1,E_2,E_3\right)U^{-1}
+{\mbox{\rm diag}}\left(A,0,0\right)
\right]\Psi,
\end{eqnarray}
where $\Psi^T\equiv(\nu_e,\nu_\mu,\nu_\tau)$ is the flavor
eigenstate, $U$ is the Pontecorvo-Maki-Nakagawa-Sakata
\cite{Pontecorvo:1957cp,Pontecorvo:1958cp,Maki:1962mu} (PMNS)
matrix,\footnote{Following S.T. Petcov \cite{p}, we call
$U$ the PMNS matrix.} and $E_j\equiv\sqrt{m_j^2
+{\vec {\relax{\kern .1em p}}}^2}$.
Throughout this paper we assume that the density of matter is constant
for simplicity.  The case of matter with slowly varying density
will be briefly discussed at the end of the paper.

Here let us consider more general case with $N$ neutrino flavors
and with general matter effect:
\begin{eqnarray}
i{d\Psi \over dt}=
\left(U{\cal E}U^{-1}+{\cal A}\right)\Psi,
\label{eqn:schn}
\end{eqnarray}
where
\begin{eqnarray}
{\cal E}&\equiv&{\mbox{\rm diag}}\left(E_1,E_2,\cdots,E_N\right)\\
{\cal A}&\equiv&{\mbox{\rm diag}}\left(A_1,A_2,\cdots,A_N\right),
\end{eqnarray}
$U$ is the $N\times N$ PMNS matrix
and $\Psi^T\equiv(\nu_{\alpha_1},\nu_{\alpha_2},\cdots,\nu_{\alpha_N})$
is the flavor eigenstate.
Without the matter effect (i.e., $A_j=0$, $j=1,\cdots,N$),
(\ref{eqn:schn}) can be easily solved and the oscillation
probability
$P(\nu_\alpha\rightarrow\nu_\beta)_{\mbox{\rm\scriptsize vacuum}}$ is
given by
\begin{eqnarray}
P(\nu_\alpha\rightarrow\nu_\beta)_{\mbox{\rm\scriptsize vacuum}}&=&
\delta_{\alpha\beta}-2\sum_{j,k}U_{\alpha j}U^\ast_{\beta j}
U^\ast_{\alpha k}U_{\beta k}\sin^2
\left({\Delta E_{jk}L \over 2}\right)\nonumber\\
&{\ }&-i\sum_{j,k}U_{\alpha j}U^\ast_{\beta j}
U^\ast_{\alpha k}U_{\beta k}\sin
\left(\Delta E_{jk}L\right),
\label{eqn:probv}
\end{eqnarray}
where $\Delta E_{jk}\equiv E_j-E_k$ and
the second and the third terms on the right hand side
correspond to CP-conserving and CP-violating probabilities, respectively.

With the nonvanishing matter effect, on the other hand,
explicit evaluation of the probability is difficult but
the $N\times N$ matrix $U{\cal E}U^{-1}+{\cal A}$ on the right
hand side of (\ref{eqn:schn}) can be formally diagonalized
by an $N\times N$ unitary matrix $U^M$:
\begin{eqnarray}
U{\cal E}U^{-1}+{\cal A}=U^M{\cal E}^M(U^M)^{-1},
\end{eqnarray}
where
\begin{eqnarray}
{\cal E}^M\equiv{\mbox{\rm diag}}\left(E^M_1,E^M_2,\cdots,E^M_N\right),
\end{eqnarray}
and $E^M_j$ stands for the eigenvalue of $U{\cal E}U^{-1}+{\cal A}$.
As in the case of the oscillation probability in vacuum,
we can formally solve (\ref{eqn:schn}) and express the oscillation
probability
$P(\nu_\alpha\rightarrow\nu_\beta)_{\mbox{\rm\scriptsize matter}}$ as
\begin{eqnarray}
P(\nu_\alpha\rightarrow\nu_\beta)_{\mbox{\rm\scriptsize matter}}&=&
\delta_{\alpha\beta}-2\sum_{j,k}U^M_{\alpha j}U^{M\ast}_{\beta j}
U^{M\ast}_{\alpha k}U^M_{\beta k}\sin^2
\left({\Delta E^M_{jk}L \over 2}\right)\nonumber\\
&{\ }&-i\sum_{j,k}U^M_{\alpha j}U^{M\ast}_{\beta j}
U^{M\ast}_{\alpha k}U^M_{\beta k}\sin
\left(\Delta E^M_{jk}L\right),
\label{eqn:probm}
\end{eqnarray}
where $\Delta E^M_{jk}\equiv E^M_j-E^M_k$ and
the second and the third terms on the right hand side
correspond to T-conserving and T-violating probabilities, respectively.

Now let us assume that $|\Delta E_{jk}L|\ll 1$ and
$|\Delta E^M_{jk}L|\ll 1$ are satisfied, where the latter
follows if $|\Delta E_{jk}L|\ll 1$ and $|A_jL|\ll 1$.
Then we can expand the sine functions in
(\ref{eqn:probv}) and (\ref{eqn:probm}).  The zeroth order term
is obviously $\delta_{\alpha\beta}$ for both probabilities.
The term linear in $\Delta E_{jk}L$ vanishes, since
\begin{eqnarray}
\sum_{j,k}U_{\alpha j}U^\ast_{\beta j}
U^\ast_{\alpha k}U_{\beta k}\Delta E_{jk}L
&=&L\sum_{j,k}U_{\alpha j}U^\ast_{\beta j}
U^\ast_{\alpha k}U_{\beta k}\left(E_j-E_k\right)\nonumber\\
&=&L\left[\delta_{\alpha\beta}\sum_{j}U_{\alpha j}U^\ast_{\beta j}E_j
-\delta_{\alpha\beta}\sum_{k}U^\ast_{\alpha k}U_{\beta k}E_k\right]
\nonumber\\
&=&L\delta_{\alpha\beta}\left[\left(U{\cal E}U^{-1}\right)_{\alpha\beta}
-\left(U{\cal E}U^{-1}\right)_{\beta\alpha}\right]\nonumber\\
&=&0,
\end{eqnarray}
where $\delta_{\alpha\beta}$ has been obtained from the
unitarity condition
$\displaystyle\sum_{j}U_{\alpha j}U^\ast_{\beta j}=\delta_{\alpha\beta}$,
and the last equality holds because the inside of
the square bracket vanishes for $\alpha=\beta$.  Similarly we have
\begin{eqnarray}
\sum_{j,k}U^M_{\alpha j}U^{M\ast}_{\beta j}
U^{M\ast}_{\alpha k}U^M_{\beta k}\Delta E^M_{jk}L
=L\delta_{\alpha\beta}\left[\left(U^M{\cal E}(U^M)^{-1}\right)_{\alpha\beta}
-\left(U^M{\cal E}(U^M)^{-1}\right)_{\beta\alpha}\right]=0.
\end{eqnarray}

The first nontrivial case is the term quadratic in
$\Delta E_{jk}L$ and $\Delta E^M_{jk}L$.
From (\ref{eqn:probm}) we have the term quadratic in
$\Delta E^M_{jk}L$ (up to a factor $-1/2$)

\begin{eqnarray}
&{\ }&\sum_{j,k}U^M_{\alpha j}U^{M\ast}_{\beta j}
U^{M\ast}_{\alpha k}U^M_{\beta k}\left(\Delta E^M_{jk}L\right)^2\nonumber\\
&=&L^2\sum_{j,k}U^M_{\alpha j}U^{M\ast}_{\beta j}
U^{M\ast}_{\alpha k}U^M_{\beta k}\left[
(E^M_j)^2-2E^M_jE^M_k+(E^M_k)^2\right]\nonumber\\
&=&L^2\left[\delta_{\alpha\beta}
\sum_{j}U^M_{\alpha j}U^{M\ast}_{\beta j}(E^M_j)^2
+\delta_{\alpha\beta}\sum_{k}U^{M\ast}_{\alpha k}U^M_{\beta k}(E^M_k)^2
-2\sum_{j}U^M_{\alpha j}U^{M\ast}_{\beta j}E^M_j
\sum_{k}U^{M\ast}_{\alpha k}U^M_{\beta k}E^M_k
\right].\nonumber\\
\label{eqn:quadratic}
\end{eqnarray}
Here we note the following properties:
\begin{eqnarray}
\sum_{j}U^M_{\alpha j}U^{M\ast}_{\beta j}E^M_j
&=&\left(U{\cal E}U^{-1}+{\cal A}\right)_{\alpha\beta}
=\left(U{\cal E}U^{-1}\right)_{\alpha\beta}
+\delta_{\alpha\beta}{\cal A}_\alpha,\\
\sum_{j}U^M_{\alpha j}U^{M\ast}_{\beta j}(E^M_j)^2
&=&\left[U^M({\cal E}^M)^2(U^M)^{-1}\right]_{\alpha\beta}
=\left\{\left[U^M{\cal E}^M(U^M)^{-1}\right]^2\right\}_{\alpha\beta}
\nonumber\\
&=&\left[\left(U{\cal E}U^{-1}+{\cal A}\right)^2\right]_{\alpha\beta}
\nonumber\\
&=&\left(U{\cal E}^2U^{-1}\right)_{\alpha\beta}
+({\cal A}_\alpha+{\cal A}_\beta)
\left(U{\cal E}U^{-1}\right)_{\alpha\beta}
+\delta_{\alpha\beta}({\cal A}_\alpha)^2.
\end{eqnarray}
Thus (\ref{eqn:quadratic}) becomes
\begin{eqnarray}
&{\ }&L^2\delta_{\alpha\beta}\left\{
\left[U^M({\cal E}^M)^2(U^M)^{-1}\right]_{\alpha\beta}
+\left[U^M({\cal E}^M)^2(U^M)^{-1}\right]_{\beta\alpha}\right\}
\nonumber\\
&{\ }&-2L^2\left[U^M{\cal E}^M(U^M)^{-1}\right]_{\alpha\beta}
\left[U^M{\cal E}^M(U^M)^{-1}\right]_{\beta\alpha}
\nonumber\\
&=&2L^2\delta_{\alpha\beta}
\left[\left(U{\cal E}^2U^{-1}\right)_{\alpha\alpha}
+2{\cal A}_\alpha
\left(U{\cal E}U^{-1}\right)_{\alpha\alpha}
+({\cal A}_\alpha)^2\right]\nonumber\\
&{\ }&-2L^2\left[\left(U{\cal E}U^{-1}\right)_{\alpha\beta}
+\delta_{\alpha\beta}{\cal A}_\alpha\right]
\left[\left(U{\cal E}U^{-1}\right)_{\beta\alpha}
+\delta_{\alpha\beta}{\cal A}_\alpha\right]\nonumber\\
&=&2L^2\left[\delta_{\alpha\beta}\left(U{\cal E}^2U^{-1}\right)_{\alpha\alpha}
-\left(U{\cal E}U^{-1}\right)_{\alpha\beta}
\left(U{\cal E}U^{-1}\right)_{\beta\alpha}\right],
\label{eqn:quadratic2}
\end{eqnarray}
where all the contributions of the matter effect
have disappeared in the last step.
Since the last expression in (\ref{eqn:quadratic2})
is the term quadratic in $\Delta E_{jk}L$ for
the probability in vacuum, we obtain
\begin{eqnarray}
\sum_{j,k}U^M_{\alpha j}U^{M\ast}_{\beta j}
U^{M\ast}_{\alpha k}U^M_{\beta k}\left(\Delta E^M_{jk}L\right)^2
=\sum_{j,k}U_{\alpha j}U^\ast_{\beta j}
U^\ast_{\alpha k}U_{\beta k}\left(\Delta E_{jk}L\right)^2.
\end{eqnarray}

Next let us turn to the term cubic in $\Delta E^M_{jk}L$.
It is given by (up to a factor $i/3!$)
\begin{eqnarray}
&{\ }&\sum_{j,k}U^M_{\alpha j}U^{M\ast}_{\beta j}
U^{M\ast}_{\alpha k}U^M_{\beta k}\left(\Delta E^M_{jk}L\right)^3\nonumber\\
&=&L^3\sum_{j,k}U^M_{\alpha j}U^{M\ast}_{\beta j}
U^{M\ast}_{\alpha k}U^M_{\beta k}\left[
(E^M_j)^3-3(E^M_j)^2E^M_k+3E^M_j(E^M_k)^2-(E^M_k)^3
\right]\nonumber\\
&=&L^3\delta_{\alpha\beta}\left\{
\left[U^M({\cal E}^M)^3(U^M)^{-1}\right]_{\alpha\beta}
-\left[U^M({\cal E}^M)^3(U^M)^{-1}\right]_{\beta\alpha}\right\}
\nonumber\\
&{\ }&-3L^3\left[U^M({\cal E}^M)^2(U^M)^{-1}\right]_{\alpha\beta}
\left[U^M{\cal E}^M(U^M)^{-1}\right]_{\beta\alpha}
\nonumber\\
&{\ }&+3L^3\left[U^M{\cal E}^M(U^M)^{-1}\right]_{\alpha\beta}
\left[U^M({\cal E}^M)^2(U^M)^{-1}\right]_{\beta\alpha}
\nonumber\\
&=&-3L^3\left[\left(U{\cal E}^2U^{-1}\right)_{\alpha\beta}
+({\cal A}_\alpha+{\cal A}_\beta)
\left(U{\cal E}U^{-1}\right)_{\alpha\beta}
+\delta_{\alpha\beta}({\cal A}_\alpha)^2
\right]
\left[\left(U{\cal E}U^{-1}\right)_{\beta\alpha}
+\delta_{\alpha\beta}{\cal A}_\alpha\right]
\nonumber\\
&{\ }&+3L^3\left[\left(U{\cal E}U^{-1}\right)_{\alpha\beta}
+\delta_{\alpha\beta}{\cal A}_\alpha\right]
\left[\left(U{\cal E}^2U^{-1}\right)_{\beta\alpha}
+({\cal A}_\alpha+{\cal A}_\beta)
\left(U{\cal E}U^{-1}\right)_{\beta\alpha}
+\delta_{\alpha\beta}({\cal A}_\alpha)^2
\right]
\nonumber\\
&=&-3L^3\left[\left(U{\cal E}^2U^{-1}\right)_{\alpha\beta}
\left(U{\cal E}U^{-1}\right)_{\beta\alpha}
-\left(U{\cal E}U^{-1}\right)_{\alpha\beta}
\left(U{\cal E}^2U^{-1}\right)_{\beta\alpha}\right],
\label{eqn:cubic}
\end{eqnarray}
where all the contributions of the matter effect
have disappeared again in the last step.
Since the last expression in (\ref{eqn:cubic})
is the term cubic in $\Delta E_{jk}L$ for
the probability in vacuum, we obtain
\begin{eqnarray}
\sum_{j,k}U^M_{\alpha j}U^{M\ast}_{\beta j}
U^{M\ast}_{\alpha k}U^M_{\beta k}\left(\Delta E^M_{jk}L\right)^3
=\sum_{j,k}U_{\alpha j}U^\ast_{\beta j}
U^\ast_{\alpha k}U_{\beta k}\left(\Delta E_{jk}L\right)^3.
\label{eqn:cubic2}
\end{eqnarray}
It turns out that the matter contributions in the terms
of ${\cal O}((\Delta E_{jk}L)^4)$ or higher are not canceled
and we have
\begin{eqnarray}
P(\nu_\alpha\rightarrow\nu_\beta)_{\mbox{\rm\scriptsize matter}}=
P(\nu_\alpha\rightarrow\nu_\beta)_{\mbox{\rm\scriptsize vacuum}}
+{\cal O}((\Delta E_{jk}L)^4).
\end{eqnarray}

We note in passing that the equation (\ref{eqn:cubic2})
gives another proof of the Harrison-Scott identity \cite{Harrison:2000df}
for the case with three flavors\footnote{A different form of the
quantity $J^M/J$ has been given in \cite{Krastev:1988yu}.}
\begin{eqnarray}
J^M\Delta E^M_{31}\Delta E^M_{32}\Delta E^M_{21}=
J\Delta E_{31}\Delta E_{32}\Delta E_{21},
\end{eqnarray}
for
\begin{eqnarray}
\sum_{j,k}U^M_{\alpha j}U^{M\ast}_{\beta j}
U^{M\ast}_{\alpha k}U^M_{\beta k}\left(\Delta E^M_{jk}\right)^3
&=&i\sum_{j<k}\Im \left(U^M_{\alpha j}U^{M\ast}_{\beta j}
U^{M\ast}_{\alpha k}U^M_{\beta k}\right)
\left(\Delta E^M_{jk}\right)^3\nonumber\\
&=&iJ^M\left[-\left(\Delta E^M_{13}\right)^3
+\left(\Delta E^M_{23}\right)^3+\left(\Delta E^M_{12}\right)^3\right]
\nonumber\\
&=&-3iJ^M\Delta E^M_{31}\Delta E^M_{32}\Delta E^M_{21}
\nonumber\\
&=&\sum_{j,k}U_{\alpha j}U^\ast_{\beta j}
U^\ast_{\alpha k}U_{\beta k}\left(\Delta E_{jk}\right)^3
\nonumber\\
&=&-3iJ\Delta E_{31}\Delta E_{32}\Delta E_{21},
\end{eqnarray}
where
\begin{eqnarray}
J^M&\equiv&\Im \left(U^M_{\alpha 1}U^{M\ast}_{\beta 1}
U^{M\ast}_{\alpha 2}U^M_{\beta 2}\right)\\
J&\equiv&\Im \left(U_{\alpha 1}U^\ast_{\beta 1}
U^\ast_{\alpha 2}U_{\beta 2}\right)
\end{eqnarray}
are the Jarlskog factors in matter and in vacuum, respectively,
and we have used the fact
$a^3+b^3+c^3=a^3+b^3-(a+b)^3=-3ab(a+b)=3abc$ for $a+b+c=0$
($a\equiv\Delta E_{13}$, $b\equiv\Delta E_{32}$,
$c\equiv\Delta E_{21}$).

For long baseline experiments such as JHF \cite{jhf}
with relatively low energy
($E_\nu\sim$ 1GeV, $L\sim$ 300km), the larger mass squared difference
$|\Delta m^2_{32}|\sim 3\times 10^{-3}$eV$^2$
gives $|\Delta m^2_{32}L/2E|\sim{\cal O}(1)$ and our assumption
does not hold.  In fact it has been shown \cite{m2} that
there is some contribution from the matter effect to CP violation
at the JHF neutrino experiment.

So far we have assumed that the density of matter is approximately
constant.  However, even if the density depends on
the position, if adiabatic treatment is allowed (i.e.,
$|dU^M/dt|\ll|E^M_j|$) then we can apply our argument
to each interval in which the density can be regarded as
approximately constant.  Hence vacuum mimicking phenomena
occur if adiabatic treatment is justified and
$|\Delta E_{jk}L|\ll 1$ and $|A_jL|\ll 1$ are
satisfied.

After this paper was submitted to the preprint archive,
the author has learned from E. Akhmedov that
the result here holds not only in matter of approximately constant density,
but also in the case of an arbitrary density profile \cite{Akhmedov:2001cs}.
The author would like to thank E. Akhmedov for useful communication,
and S.T. Petcov and C.~Pe\~na-Garay for discussions.  This
research was supported in part by a Grant-in-Aid for Scientific
Research of the Ministry of Education, Science and Culture,
\#12047222, \#13640295.

\end{document}